\documentclass[sigconf,sigconf]{acmart}

\usepackage[utf8]{inputenc}
\usepackage{xcolor}
\usepackage{graphicx}
\usepackage{soul}
\usepackage{makecell}
\usepackage{caption}

\AtBeginDocument{%
    \providecommand\BibTeX{{%
    \normalfont B\kern-0.5em{\scshape i\kern-0.25em b}\kern-0.8em\TeX}}}

\setcopyright{acmlicensed}
\copyrightyear{2025}
\acmYear{2025}
\setcopyright{cc}
\setcctype{by}
\acmConference[CHI '25]{CHI Conference on Human Factors in Computing Systems}{April 26-May 1, 2025}{Yokohama, Japan}
\acmBooktitle{CHI Conference on Human Factors in Computing Systems (CHI '25), April 26-May 1, 2025, Yokohama, Japan}\acmDOI{10.1145/3706598.3713315}
\acmISBN{979-8-4007-1394-1/25/04}

\setcopyright{cc}
\setcctype{by}
\copyrightyear{2025}
\acmYear{2025}
\acmDOI{10.1145/3706598.3713315}
\acmISBN{979-8-4007-1394-1/25/04}
\acmConference[CHI '25]{CHI Conference on Human Factors in Computing Systems}{April 26--May 01, 2025}{Yokohama, Japan}
\acmBooktitle{CHI Conference on Human Factors in Computing Systems (CHI '25), April 26--May 01, 2025, Yokohama, Japan}

\definecolor{green}{HTML}{005B5A}
\definecolor{blue}{HTML}{002753}
\definecolor{orange}{HTML}{D64900}
\definecolor{red}{HTML}{CB0000}

\begin{document}

\title[The Odyssey Journey]{The Odyssey Journey: Top-Tier Medical Resource Seeking for Specialized Disorder in China}

\author{Ka I Chan}
\orcid{0009-0001-4560-1702}
\affiliation{\institution{Department of Computer Science and Technology, Beijing National Research Center for Information Science and Technology, Global Innovation Exchange (GIX) Institute \\ Tsinghua University}
\city{Beijing}
\country{China}}
\email{ckinicola@gmail.com}

\author{Siying Hu}
\orcid{0000-0002-3824-2801}
\affiliation{\institution{Department of Computer Science \\ City University of Hong Kong}
\city{Hong Kong}
\country{China}}
\email{siyinghu-c@my.cityu.edu.hk}

\author{Yuntao Wang}
\authornote{corresponding author}
\orcid{0000-0002-4249-8893}
\affiliation{\institution{Key Laboratory of Pervasive Computing, Ministry of Education, Department of Computer Science and Technology \\ Tsinghua University}
\city{Beijing}
\country{China}}
\affiliation{\institution{School of Computer Science and Technology \\ Qinghai University}
\city{Xining}
\country{China}}
\email{yuntaowang@tsinghua.edu.cn}

\author{Xuhai Xu}
\orcid{0000-0001-5930-3899}
\affiliation{\institution{Google}
\city{New York City}
\state{New York}
\country{USA}}
\email{orson@google.com}

\author{Zhicong Lu}
\orcid{0000-0002-7761-6351}
\affiliation{\institution{Department of Computer Science \\ George Mason University}
\city{Fairfax}
\state{Virginia}
\country{USA}}
\email{zlu6@gmu.edu}

\author{Yuanchun Shi}
\orcid{0000-0003-2273-6927}
\affiliation{\institution{Department of Computer Science and Technology, Beijing National Research Center for Information Science and Technology \\ Tsinghua University}
\city{Beijing}
\country{China}}
\affiliation{\institution{Qinghai University}
\city{Xining}
\country{China}}
\email{shiyc@tsinghua.edu.cn}

\renewcommand{\shortauthors}{Chan, et al.}
\begin{abstract}
It is pivotal for patients to receive accurate health information, diagnoses, and timely treatments. However, in China, the significant imbalanced doctor-to-patient ratio intensifies the information and power asymmetries in doctor-patient relationships. Health information-seeking, which enables patients to collect information from sources beyond doctors, is a potential approach to mitigate these asymmetries. While HCI research predominantly focuses on common chronic conditions, our study focuses on specialized disorders, which are often familiar to specialists but not to general practitioners and the public.
With Hemifacial Spasm (HFS) as an example, we aim to understand patients' health information and top-tier\footnote{Top-tier medical resources and services refer to medical institutions that are well-known for their specialized departments and are typically located in big cities. These institutions have highly qualified medical professionals and are equipped with state-of-the-art facilities and equipment. In this paper, this mainly refers to hospitals renowned for their expertise in treating hemifacial spasm.} medical resource seeking journeys in China. Through interviews with three neurosurgeons and 12 HFS patients from rural and urban areas, and applying Actor-Network Theory, we provide empirical insights into the roles, interactions, and workflows of various actors in the health information-seeking network. We also identified five strategies patients adopted to mitigate asymmetries and access top-tier medical resources, illustrating these strategies as subnetworks within the broader health information-seeking network and outlining their advantages and challenges.
\end{abstract}


\maketitle

\begin{figure*}
    \centering\includegraphics[width=\textwidth]{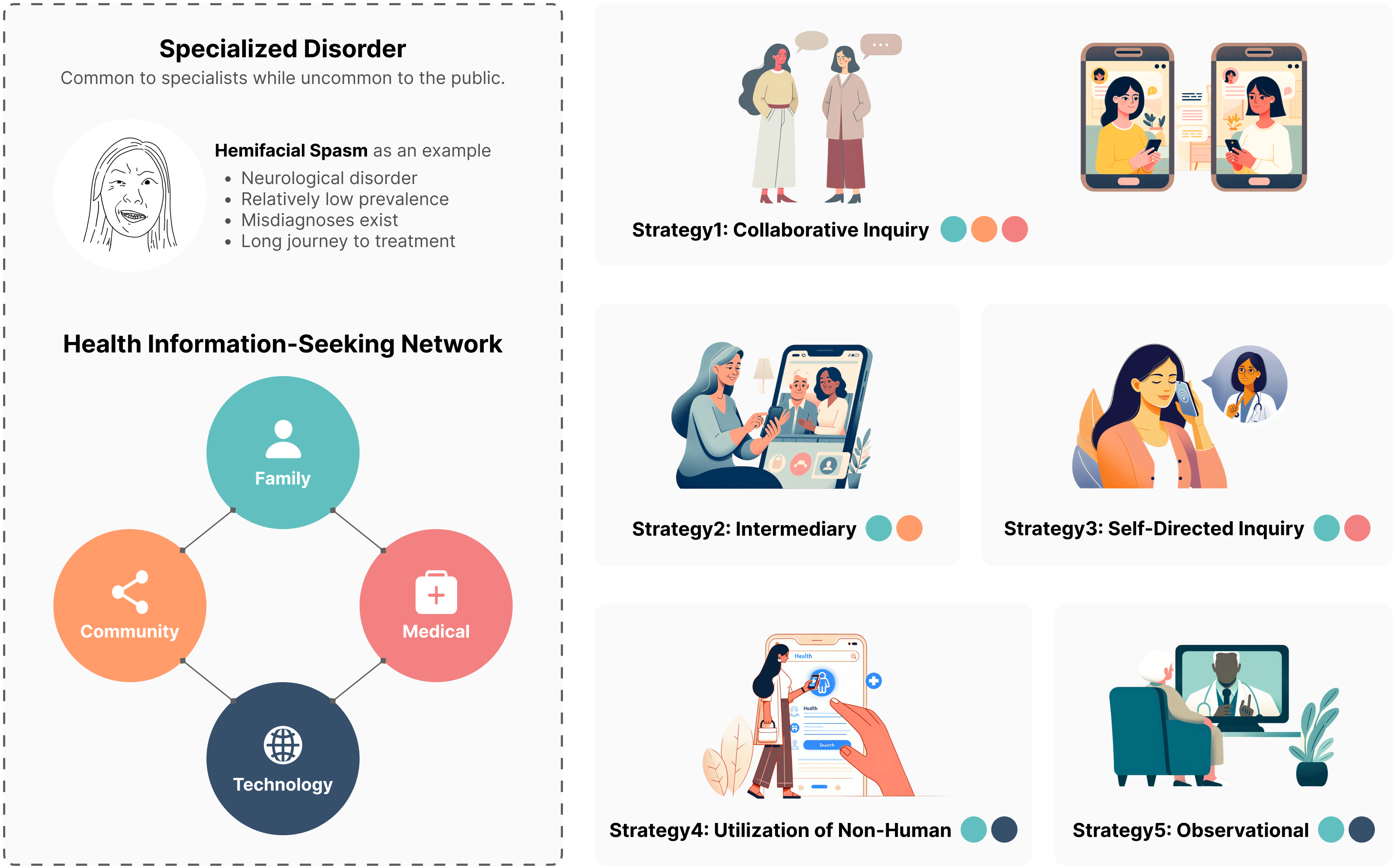}
    \caption{\textbf{Study Overview.} In this paper, we focus on specialized disorders, taking Hemifacial Spasm as an example to better understand how patients mitigate the information and power asymmetries in the doctor-patient relationship through health information-seeking in China. We proposed a health information-seeking network formed by four units and identified five strategies, each involving different units, utilized by patients to access top-tier medical resources.}
  \label{fig:teaser}
  \Description{Figure 1 presents the overview of our study and is divided into three sections. The first section, located at the top left, outlines that specialized disorders are common to specialists but uncommon to the public. Hemifacial Spasm is used as an example in this paper; its characteristics are listed, such as being a neurological disorder, having relatively low prevalence, existing misdiagnoses, and a long journey to treatment, and it includes an image of a woman's face twitching. The second section, at the bottom left, includes a network diagram with four interconnected nodes labeled Family, Medical, Technology, and Community. Each node is represented by a different color and icon, indicating the various sources from which patients might seek health information. The third section, on the right, illustrates five different strategies used by patients to access top-tier medical resources. Strategy 1: Collaborative Inquiry is depicted by two women conversing, discussing medical issues or sharing information. Strategy 2: Intermediary shows a woman using a tablet to facilitate a video call between an elderly woman and a healthcare provider, suggesting the use of intermediaries in obtaining medical advice. Strategy 3: Self-Directed Inquiry is illustrated by a woman using her smartphone to consult with a doctor she is familiar with. Strategy 4: Utilization of Non-Human features a woman using her smartphone, likely searching for health information independently. Strategy 5: Observational depicts a woman watching a health-related program on a television.}
\end{figure*}


\section{Introduction}
\label{sec:intro}

In the context of patients seeking medical assistance, doctors often play a crucial and authoritative role \cite{cutilli2010seeking}, as they are generally regarded as common and reliable sources of accurate information and adequate treatment options. However, in China, the doctor-to-patient ratio is highly imbalanced\footnote{China has only 1.98 medical doctors per 1,000 of the population, including generalist and specialist medical practitioners, falling short of the World Health Organization's standard for primary healthcare needs, which is more than 2.3 health workers per 1,000 \cite{whoMedicalDoctors}.}, thus accessing top-tier medical professionals can be challenging for many patients. Often, they can only consult with doctors within their reach (who may not have the right expertise) and have to rely on the authority and health information those doctors provide, without the ability to verify it due to limited information sources. This scenario contributes to the information and power asymmetries in the doctor-patient relationship.

Individuals who become aware of abnormalities in their bodies typically seek and utilize various health information resources, such as healthcare systems, fellow patients, friends, family, and the internet. As such, actively seeking health information is a vital approach to mitigate the potential information and power asymmetries in the doctor-patient relationship, empowering patients with knowledge beyond those provided by their doctors. The use of online sources for health information-seeking has become prevalent, with patients using search engines and social media for self-diagnosis and exploring potential treatments \cite{de2014seeking, wang2012using}. It can lead to significant offline impacts, affecting patients' emotions and influencing their health-related decisions \cite{chi2020connections, kuhlthau1991inside, lambert2007health, ogan2008embedding, fox2002vital}.

Extensive research within the HCI communities has been dedicated to understanding the practice and design of tools supporting health information work in the context of common chronic conditions, such as diabetes, dementia, cardiovascular disease \cite{desai2019personal, maddali2022investigating, 10.1145/3613904.3642701}, which are generally well-known to the public. In contrast, studies focusing on rare diseases \cite{macleod2015rare, macleod2017grateful, nielsen2023patient} and specific disorders \cite{sannon2019really, mckillop2018designing, pichon2021divided} are fewer. 
There is a significant gap in understanding conditions that are familiar to specialized doctors but not to the general public. These conditions typically have relatively low prevalence, which means that even general practitioners may not be familiar with them. Thus, patients are often struggling to reach specific departments for accurate diagnosis. In this paper, we refer to these conditions as \textbf{specialized disorders} and use Hemifacial Spasm (HFS) as an example to better understand this challenge.

Unlike common chronic conditions, HFS as a specialized disorder has a total prevalence of only 9.8 per 100,000 \cite{nilsen2004prevalence} and is predominantly managed within the fields of neurosurgery, leading to limited public exposure and access to relevant health information. The uncommonness of HFS, coupled with its similarities to facial paralysis, increases the risk of misdiagnosis and misinformation, making it challenging for patients to find adequate and satisfactory treatment options. This issue is particularly acute in rural areas, where lower health literacy levels may impede individuals' ability to obtain proper health information \cite{aljassim2020health}. Consequently, these factors exacerbate the gap between doctors' authoritative and professional knowledge and patients' limited information sources, intensifying the information and power asymmetries in doctor-patient relationships.

HFS patients often undergo a long journey before they can access medical treatments, averaging 11.4 ± 8.5 years \cite{wang1998hemifacial}. Our research aims to understand how these patients are empowered by leveraging various approaches, such as actively seeking information, utilizing the internet, and engaging in social networking. This empowerment is critical in overcoming the information and power asymmetries that exist in doctor-patient relationships, thereby improving their access to health information and medical resources, particularly to top-tier medical services. It is significant to explore how HFS patients navigate the needed health information, gain access to effective treatment options and channels leading to top-tier medical resources, thereby our study aim to investigate the following research questions: 
\begin{itemize}
    \item[RQ1:] What are the primary sources from which HFS patients seek health and healthcare information?
    \item[RQ2:] What strategies do HFS patients adopt to acquire top-tier medical resources and services, and how do these strategies empower them to mitigate the power and information asymmetries in the doctor-patient relationship?
\end{itemize}

This study employs qualitative research methods to explore the health information and top-tier medical resource seeking journey of HFS patients in China. We conducted semi-structured interviews with three neurosurgeons and 12 HFS patients from rural (N=6) and urban areas (N=6) of China. 
During the first round of data analysis using thematic analysis, we discovered that health and medical information seeking for HFS patients mainly relies on guanxi\footnote{Guanxi refers to the complex dynamics of social networks and relationships in Chinese society, which are presonalized, subjectively perceived as close, and rich in resources \cite{bian2019guanxi}.} and social interactions, especially for those from rural areas, with technology playing a weak role despite its great potential to assist patients. Therefore, we applied Actor-Network Theory (ANT) in our second round of data analysis, which enabled us to identify and comprehend the various actors involved, analyzing and discussing both human and non-human entities at the same level. This approach helped us uncover how these actors collaborate and interact, thereby forming stable and effective health information channels. Furthermore, we observed the various strategies employed within this health information-seeking network. This analytical approach highlights how social networks and technology can be leveraged to diversify and strengthen information channels, thereby providing patients with broader and more effective access to various treatment resources.

We contribute to the HCI communities by building upon existing research on health information-seeking in a broader context. With HFS as a research lens, our research introduces a novel information-seeking perspective by focusing on specialized disorders that are uncommon among general patients and health practitioners. This particular focus sheds light on the exacerbated information and power asymmetries in doctor-patient relationships, allowing us to delve deeper into the strategies that patients utilize to access top-tier medical resources. Our contributions include: 
\begin{itemize}
    \item[1.] We provided empirical evidence and employ Actor-Network Theory to elucidate the roles, interactions, and workflows of various actors within the HFS patients' health information-seeking network.
    \item[2.] We identified five strategies utilized by patients with specialized disorders from both urban and rural areas in China to access top-tier medical resources and services. We illustrated these strategies as subnetworks within the broader health information-seeking network. Additionally, we detailed the advantages and challenges associated with each of these strategies.
    \item[3.] We found that guanxi and strong ties, as parts of human infrastructure, plays crucial roles in the process of seeking top-tier medical resources in China. Additionally, the potential of technology to empower patients assists in mitgating the challenges and asymmetries in medical access.
\end{itemize}


\section{Background and Related Work}
\label{sec:related_work}

\subsection{Hemifacial Spasm}
\label{subsec:hemifacial_spasm}
Hemifacial Spasm (HFS) is a chronic neurological disorder characterized by unilateral tonic and clonic contractions of facial muscles innervated by the facial nerve \cite{tan2002hemifacial}, often originating from the orbicularis oculi muscle \cite{wang1998hemifacial}. The annual incidence rate of HFS is 0.78 per 100,000 \cite{auger1990hemifacial}, and the total prevalence is 9.8 per 100,000, although this may be underestimated due to misdiagnosis and lack of referral from general practitioner \cite{nilsen2004prevalence}. HFS can significantly impact a patient's quality of life, leading to social discomfort and awkwardness, emotional distress, and functional limitations \cite{reimer2005health}. HFS can be accentuated during speaking and eating, as well as psychological tension \cite{abbruzzese2011hemifacial, wilkins1991hemifacial}. It is important to note that attributing these symptoms to stress and anxiety can potentially lead to underdiagnosis of HFS. 

The treatment options for HFS include drug therapy, Botox injections, acupuncture, and microvascular decompression (MVD) surgery. Medications, Botox injections, and acupuncture offer limited efficacy and temporary relife, serving as low-risk palliative treatments \cite{wang1998hemifacial, wang2012effectiveness}. Among these, MVD is the only radical treatment with a high success rate for addressing the underlying cause of HFS \cite{montava2016long}. However, MVD carries inherent risks and potential complications, with approximately one-third of patients (33\%±8\%) experiencing a delayed cure in post-operation \cite{sindou2018microvascular}. Additionally, early surgical intervention is preferred because long-term HFS patients tend to have significantly poorer postoperative outcomes \cite{rosenstengel2012hemifacial}. Notably, the primary treatments employed in China are Botox injections and acupuncture \cite{wang2014clinical}. The shared decision-making process is crucial when choosing between surgery and non-surgery options \cite{ubbink2015shared}. This process involves healthcare providers introducing the benefits and potential harms of available treatment options to patients, weighing the probabilities of outcomes alongside the patients' values and needs to reach a consensus on the decision \cite{frosch1999shared}. The optimal treatment is the one that best aligns with the patient's preferences, taking into account the pros and cons of each option \cite{barry2012shared}. Consequently, patients who have high anesthetic or surgical risks may reject surgery \cite{rosenstengel2012hemifacial}.

\subsection{Doctor-Patient Relationships in Chinese Context}
The doctor-patient relationship is pivotal in ensuring patients receive accurate health information, diagnoses, and timely treatments. However, China's situation is distinct, especially when compared to developed countries like the U.S., the U.K., and Japan. These high-income countries typically have around 10 nurses and 3 doctors per 1000 people, while China has only about 1.3 nurses and 1.7 doctors \cite{qin2013too}. 
This shortage of healthcare professionals in China may be attributed to the relatively unattractive income, unlike in developed western countries where being a doctor is a high-income occupation with significant returns on professional education \cite{nicholson2008medical}.

In China, as society progresses, people are becoming more conscious of their health, leading to increased demand for medical resources. However, this demand is met with challenges such as power and information asymmetries, along with unequal allocation of medical resources between rural and urban areas, intensifying the tension in doctor-patient relationships \cite{zhu2019addressing}. Most developing countries, characterized by high power distance, often place doctors in a dominant position during consultations \cite{goodyear2001power}. Patients may feel it is not their place to question their health or express their feelings, leading to brief consultations dominated by the doctor's decisions, with little room for unexpected information exchange \cite{jacobs2019relationship}. 
This dynamic contributes to the power asymmetry in doctor-patient relationships, with patients hesitant to challenge the authority of doctors. Moreover, doctors dominate the consultation in terms of knowledge, skills, time, and resources, which exacerbates the knowledge gap and information asymmetry \cite{zhang2023disruption}. \citet{cheng2015guanxi} found that doctor-patient relationships could potentially change how doctors diagnose and treat patients, including issues such as over-examination, over-diagnosis, and over-treatment \cite{zhang2021seeking}. As a result, doctor-patient relationships in China tend towards paternalism, with patients striving for better medical resources to receive an optimal medical decision about their health.

\subsection{Health Information Seeking}
Health information-seeking refers to the methods individuals use to acquire information related to health, illness, health promotion, and health risks \cite{lambert2007health}. This process is dynamic; as individuals encounter new health information, their understanding evolves, impacting subsequent information-seeking and decision-making \cite{kleinman1978culture}. Despite the abundance of health information, many people suffer severe complications or lose their lives due to a lack of knowledge about prevention methods. This can be attributed to insufficient proactiveness, reliance on passive information acquisition, the influence of personal characteristics and information sources, as well as other contextual factors beyond patients' control \cite{lalazaryan2014review}. Therefore, exploring patients' information-seeking behaviors can provide valuable insights into strategies for effectively conveying information, enhancing self-management, and preventing disorder progression.

Individuals access health information through various channels, including healthcare systems, friends, family, and the internet. Increasingly, patients, caregivers, and the general population turn to the internet for health information \cite{fuertes2007physician}, which can significantly influence their beliefs and actions \cite{barker2008electronic, davison2013virtual}. Online platforms such as search engines, social media, and online communities offer privacy, real-time interaction, and archived health information, facilitating patient knowledge and self-diagnosis. However, barriers to online health information-seeking include low health literacy, limited information retrieval skills, unreliable information, platform censorship, and misinformation \cite{jia2021online, augustaitis2021online}.
Collaboration in information work is crucial for helping patients understand their health conditions and access healthcare resources effectively. Patients often collaborate with medical practitioners, family, friends, and fellow patients \cite{burgess2022just, klasnja2010blowing}. Health information-seeking is beneficial not only for acquiring knowledge but also for navigating the healthcare system, especially during critical circumstances such as emergency visits \cite{gui2018navigating, park2017patient}. 

In HCI, health information-seeking has been examined in various populations, including teenagers, transgender individuals, and patients with mental health issues, chronic conditions, and rare diseases \cite{dewan2024teen, augustaitis2021online, milton2024seeking, klasnja2010blowing, macleod2015rare}. \citet{gui2018navigating} elucidated the health service navigation process of parents for their children in the U.S. healthcare system and their infrastructuring work for accessing healthcare services \cite{gui2019making}. Design studies have also aimed at benefiting Black older adults and fat people in their health information-seeking efforts \cite{harrington2023trust, jonas2024we}. Notably, \citet{macleod2017grateful} explored the support needs met or neglected by different sources for patients with rare diseases and proposed technology designs to improve information sharing and navigation for these patients \cite{macleod2015rare}.

\subsection{Actor-Network Theory}
Actor-Network Theory (ANT), developed in the 1980s by Michel Callon, John Law, and Bruno Latour \cite{law1986power, law1999actor, latour2011drawing}, is a theoretical and methodological approach to social theory. ANT posits that both human and non-human entities, regarded as actors, influence the activity of a sociotechnical system \cite{crawford2020actor}. It contends that no external social force exists outside these relationships, and phenomena do not exist independently. Thus, ANT focuses on "describing" rather than "explaining" social phenomena \cite{latour2007reassembling}. It challenges traditional methods by defining non-humans as actors equal to humans. Initially, ANT explored how knowledge growth and structure could be analyzed through actor interactions \cite{law1984science}, offering a novel perspective for practical application. Despite facing both embrace and critique, ANT remains a valuable form of inquiry in various fields, such as organizational analysis, informatics, and health studies \cite{whittle2008actor}.

ANT has been effectively utilized within HCI communities. It has informed studies in participatory design \cite{gartner1996mapping, de2004construction}, the role of design materials \cite{tholander2012understanding, fuchsberger2013materials}, the development of cyberstructures \cite{randall2015creating}, mobile media use in India \cite{kumar2013mobile}, parenting practices among Latino immigrants \cite{wong2019parenting}, asymmetries in the gig economy \cite{kinder2019gig}, and care for breast cancer survivors \cite{lefkowitz2022black}. These diverse applications demonstrate ANT's effectiveness in understanding the creation and maintenance of various sociotechnical systems.

In ANT, non-human entities—such as objects, transportation devices, and texts—are valued equally with human actors \cite{sayes2014actor}. ANT provides a uniform framework "regardless of the unit of analysis" \cite{ciborra2000control}, promoting "analytical impartiality" in dealing with both human and non-human entities. This approach allows for an equal analysis of the myriad roles within the network \cite{thebault2017toward}.
As we explore the information flow and strategies in a network of individuals and resources navigating pathways to top-tier medical services, ANT enables a systematic analysis of this sociotechnical system. By balancing the importance of social networking and technology in the health information-seeking network, ANT helps us delve into the tensions and alliances between patients, doctors, social networks, and technology. This comprehensive actor-network analysis aids in understanding the strategies patients use to mitigate asymmetries in doctor-patient relationships and explores the connections and interactions between actors.


\section{Methods}
\label{sec:method}
To better understand the health information-seeking experiences and practices among HFS patients and explore our research questions, we referred to the methodologies in \citet{gui2019making} and \citet{lefkowitz2022black} for conducting interviews and qualitative analysis. Semi-structured interviews were conducted with three neurosurgeons and 12 patients in China. 

\subsection{Interviews and Recruitment}
\label{subsec:interviews}
All interviews were conducted in Mandarin and audio-recorded with participants' consent. The recordings were transcribed by the first author, a native Mandarin speaker who conducted all the interviews. Participants were fully briefed on the study's purpose and received monetary compensation after data collection. This study received approval from the university's Institutional Review Board (IRB).

\textbf{\textit{Doctor-Centered Interview.}}
Semi-structured interviews were initially conducted with three neurosurgeons (denoted as D1-D3). These neurosurgeons, with at least five years of professional experience, work in a top-tier hospital in Beijing (a first-tier city in China). The interviews were designed to gather information from a professional perspective and to gain an overview of the patients' situations. Topics discussed included HFS patients' general circumstances, their experiences within the healthcare process, and the potential challenges they might face, as seen from the professional viewpoint. All interviews were conducted remotely via audio calls in July 2023, lasted from 0.5 to 1 hour. 

\textbf{\textit{Patient-Centered Interview.}}
We then conducted semi-structured interviews with 12 individuals diagnosed with HFS from both rural and urban areas (denoted as P1-P12, Table \ref{tab:patients}). Among them, P1-P11 had undergone microvascular decompression (MVD) surgery, the only existing radical treatment for HFS, at a top-tier hospital in Beijing (a first-tier city in China). P12, despite having sufficient information and access to treatment, was hesitant about undergoing surgery. Participants living in cities across China were categorized as urban, while those in counties and countryside areas were categorized as rural. All interviews were conducted remotely using audio calls via mobile phones in August 2023, except for P2 and P3, whose interviews were conducted via text and voice messages on WeChat\footnote{A Chinese instant messaging app.} due to P2’s limited Mandarin proficiency and P3’s Hyperacusis\footnote{Hyperacusis is a condition that increases sensitivity to sound and reduces tolerance for ambient noise, making phone calls intolerable \cite{tyler2014review}.}. All patients own a smartphone except P8, who owns a basic mobile phone.

The patient interviews aimed to understand their experiences and practices related to HFS. The first part focused on how patients realized and recognized their condition, with questions like "\textit{When did you become aware the abnormalities?}" and "\textit{How did you realize that you have Hemifacial Spasm?}". The second part examined how patients sought related health and healthcare information, with questions like "\textit{Have you tried to self-diagnose? If so, how did you do that?}", "\textit{What approach did you adopt to search for health and healthcare information, and how did you search for it?}", and "\textit{How did you seek medical help at that time?}".
During each phase, the first author listened attentively without interruptions and probed further with questions such as, "\textit{Is there anything else you would like to add?}", "\textit{How did it begin?}", and "\textit{What happened before/after?}" until the participants had nothing more to add. Each interview lasted between 30 minutes to 1.5 hours.

\begin{table*}
\centering 
\caption{Summary of patients interviewed.}
\label{tab:patients}
\Description{Table 1 presents detailed information of the 12 patient participants in the semi-structured interviews, illustrating each participant's ID, sex, age, duration of Hemifacial Spasm, whether the patient has undergone MVD surgery, location, education level, and occupation.}
\begin{tabular}{ccccccc}
\hline
\multicolumn{1}{l}{\textbf{ID}} & \multicolumn{1}{l}{\textbf{Sex/Age}}  & \multicolumn{1}{l}{\textbf{HFS Duration}} & \textbf{\makecell{Undergone \\ MVD surgery}}
&\makecell{\textbf{Location} \\ (City/County, Province)}& \textbf{Education level}& \textbf{Occupation}  \\ \hline
P1 & F/47 & 8 years   &
Yes
&Urban (Jining, Shandong)& Bachelor & Engineer  \\
P2 & F/52 & 2 years   &
Yes
&Urban (Yantai, Shandong)& Middle school & Retiree  \\
P3 & F/70 & 8 years   &
Yes
&Urban (Beijing)& Associate & Retiree  \\
P4 & M/55 & 6 years   &
Yes
&Rural (Gushi, Henan)& Elementary school & Restaurant owner  \\
P5 & F/54 & 11 years   &
Yes
&Urban (Wuhan, Hubei)& Technical secondary school & Retiree  \\
P6 & M/48 & 1.5 years   &
Yes
&Rural (Wuzhai, Shanxi)& Bachelor & Government employee  \\
P7 & F/54 & 9 months   &
Yes
&Rural (Changling, Hebei)& High school & Retiree  \\
P8 & F/70 & 3 years   &
Yes
&Rural (Xiajin, Shandong)& No formal education & Farmer  \\
P9 & M/31 & 2.5 years   &
Yes
&Rural (Handan, Hebei)& Elementary school & Farmer  \\
P10 & F/37 & 2 years   &
Yes
&Urban (Luohe, Henan)& Associate & Worker  \\
P11 & F/63 & 15 years   &
Yes&Urban (Beijing)& High school & Retiree  \\
P12 & M/50 & 4 years   &
No&Rural (Shenchi, Shanxi)& Bachelor & Government employee  \\ \hline
\end{tabular}
\end{table*}

Participants were recruited through direct contacts and snowball sampling. During recruitment, they were provided with comprehensive details about the study’s design, duration, and compensation. Recognizing the sensitivity of some interview questions, we informed participants that they had complete freedom to choose what aspects of their communication to report. However, we still encountered several challenges. Many patients expressed skepticism about our intentions, and some were reluctant to share their experiences due to the traumatic nature of their illness and postoperative complications. Additionally, some rural patients who could not speak Mandarin and had low literacy refused to be interviewed. Others declined participation due to work-related time constraints. These factors contributed to initial refusals. For those who expressed interest, we communicated further and conducted interviews through WeChat. Despite our efforts, scheduling difficulties led to further refusals. Although these challenges constrained our ability to recruit a larger number of participants, those recruited were diverse in terms of locations, educational levels, occupations, and socio-economic statuses.

\subsection{Data Collection and Analysis}
\label{subsec:data_analysis}
Interview data was collected from two groups: doctors and patients. All interviews were recorded and transcribed into text for analysis. The transcripts were analyzed in two rounds using thematic analysis and Actor-Network Theory (ANT) as the analytical framework.

In the first round of thematic analysis \cite{braun2006using}, two native Mandarin-speaking researchers familiarized themselves with the transcripts and developed initial analytic interests in the HFS patients' information sources and practices. Each researcher immersed themselves in the data, marking ideas related to the research questions. They independently organized the specific language and concepts described by participants into themes through descriptive and interpretive analysis. Descriptive analysis focused on the explicit content of the transcripts, addressing research questions aimed at identifying and categorizing the information sources and strategies adopted by HFS patients. Interpretive analysis, on the other hand, sought to uncover the underlying motivations and factors influencing patients' practices and strategy choices, aiming to gain insights into their experiences and behaviors within their social contexts \cite{elliott2005descriptive}. This approach provided a deeper understanding of the motivations behind their strategy choices and how these strategies evolved over time. The researchers then generated individual lists of initial codes. Finally, they met to compare and combine the initial code lists, reaching a consensus.

ANT guided the second round of coding.
ANT is an effective approach for understanding complex sociotechnical systems. It considers how both human and non-human entities equally influence situations and analyzes how these entities interact to facilitate the flow and development of knowledge.
We adopted the actor-network research framework developed by \citet{kraal2011actor}, grounded in Callon's sociology of translation and incorporating terminology from both Callon and Latour \cite{callon1984some}. This framework interprets Callon's "moments of translation" into four steps: "interessement", "enrollment", "points of passage", and the "trial of strength". In the interessment step, actors and actants become interested in joining the network. During the enrollment step, the roles of these actors and actants are defined, assigning a specific role to play. Subsequently, one or more actors may attempt to direct, control, and manage the other actors in the points of passage step. In the trial of strength step, the stability of the network is assessed by examining whether all actors adhere to their assigned roles, ensuring the arrangement of actors is maintained. Guided by these steps, two researchers identified the actors in the health information-seeking network. They sorted the initial code list based on these actors and defined the roles of various actors in either facilitating or hindering patients' access to information, as well as the circulation of health information among these connections. This comprehensive examination allowed us to identify the strategies HFS patients adopted to access top-tier medical resources and services. All codes were translated into English and discussed by the research team to reach a consensus.

\begin{figure*}
  \includegraphics[width=\textwidth]{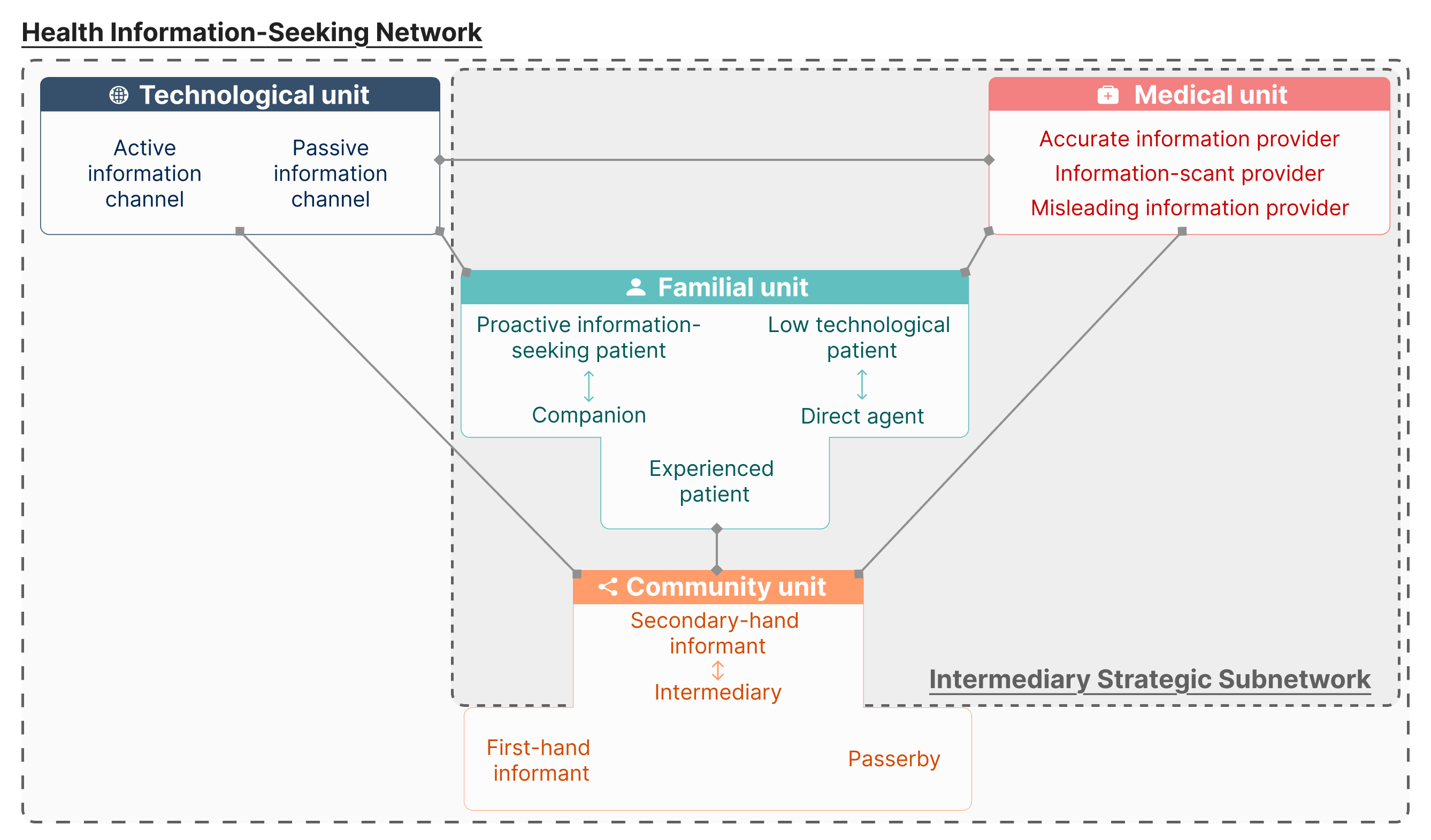}
  \caption{\textbf{ANT Graphical Depiction.} By incorporating Actor-Network Theory (ANT) during the data analysis process, we identified the main actors and roles in the health information-seeking network. We categorized these roles into four units, as shown in the figure. Additionally, we identified five strategies that form distinct strategic subnetworks within the broader health information-seeking network. The subnetwork of the intermediary strategy is depicted in the figure, and a specific example can be seen in Fig.~\ref{fig:ANT-2}.}
  \label{fig:ANT-1}
  \Description{Figure 2 illustrates a health information-seeking network structured into four main units: the technological unit, familial unit, community unit, and medical unit. The technological unit consists of two roles: active information channels and passive information channels. The familial unit includes five roles: proactive information-seeking patients, companions, low technological patients, direct agents, and experienced patients. The community unit encompasses four roles: first-hand informants, secondary-hand informants, intermediaries, and passersby. Additionally, the medical unit includes three roles: accurate information providers, information-scarce providers, and misleading information providers. These roles are all included in the broader health information-seeking network. In the intermediary strategic subnetwork, roles from the familial unit and medical unit, as well as secondary-hand informants and intermediaries from the community unit, are included.}
\end{figure*}


\section{Findings}
\label{sec:findings}
Through the qualitative data analysis, we identified the key actors within the health information-seeking network. Furthermore, this network is delineated by connections that trace the pathway of health information-seeking, ranging from initial illness awareness to the pursuit of top-tier medical resources among HFS patients. 

\begin{figure*}
  \includegraphics[width=\textwidth]{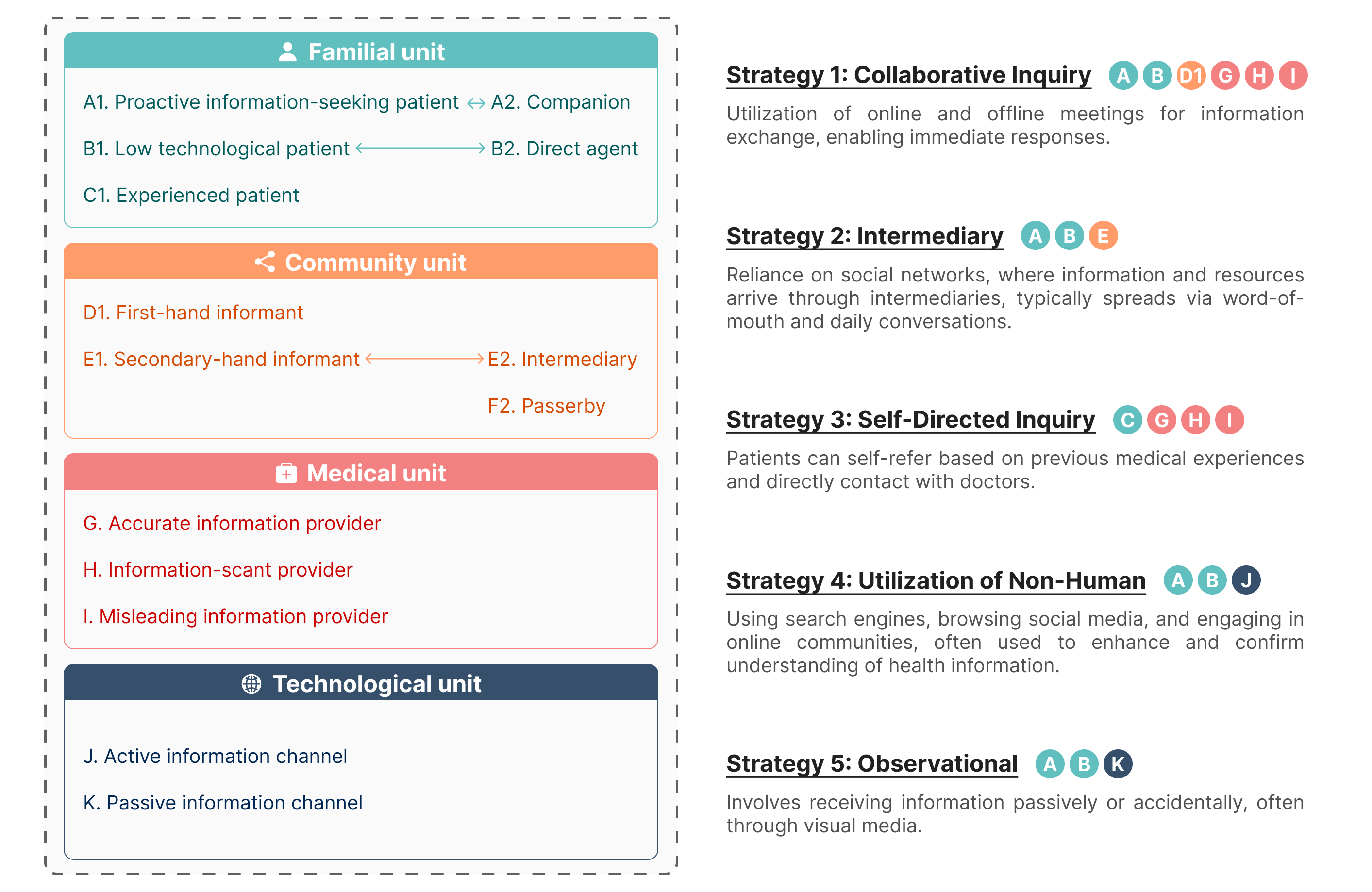}
  \caption{\textbf{Key Findings Overview.} Using Actor-Network Theory, we identified four \textbf{units}, each including various roles within the health information-seeking network of Hemifacial Spasm patients. Additionally, some of these units feature internal interactions (as shown on the left). We also identified five \textbf{strategies} for seeking top-tier medical resources of patients with specialized disorders. Each strategy involves a unique combination of units and roles, as indicated by the letter written next to the strategy name.}\endgraf {\rm This forms distinct strategic subnetworks within the broader health information-seeking network (as shown on the right).}
  \label{fig:figure2}  
  \Description{Figure 3 provides a detailed depiction of the health information-seeking network utilized by patients with specialized disorders, organized into four primary units: familial, community, medical, and technological. Each unit is represented in a distinct rectangle on the left side of the figure, detailing the specific actors and roles involved. The Familial unit includes three key interactions: (A) a proactive information-seeking patient who engages with a companion, (B) a low technological patient who interacts with a direct agent, and (C) an experienced patient who navigates the health system based on past experiences. The Community unit is composed of (D) first-hand informants who provide direct observations or experiences, (E) secondary-hand informants who interact with intermediaries to relay information, and (F) passersby who contribute sporadically to the information flow. The Medical unit consists of (G) accurate information providers who offer reliable and verified medical data, (H) information-scant providers who have limited data availability, and (I) misleading information providers who may disseminate incorrect or deceptive health information. The Technological unit is divided into (J) active information channels, and (K) passive information channels. On the right side of the figure, are five strategic approaches to top-tier medical resource seeking: 1. Collaborative Inquiry Strategy involves the utilization of both online and offline meetings to exchange information, enabling immediate responses. 2. Intermediary Strategy emphasizes reliance on social networks, where information and resources are disseminated through intermediaries, typically spreading via word-of-mouth and daily conversations. 3. Self-Directed Inquiry Strategy allows patients to self-refer based on previous medical experiences and directly contact doctors. 4. Utilization of Non-Human Strategy includes using search engines, browsing social media, and engaging in online communities to enhance and confirm the understanding of health information. 5. Observational Strategy involves receiving information passively or accidentally, often through visual media.}
\end{figure*}

\subsection{United-Actors and their Interactions in Health Information-Seeking Network}
An actor-network consists of a variety of heterogeneous actors who come together to form a network, based on connections and interactions that align with mutual interests. Our actor-network represents the workflow of health information disseminated through a diverse set of channels. We have categorized the actors within this network into distinct units, based on their nature and similarities. These units are identified as the \textcolor{green}{\textit{Familial Unit}}, the \textcolor{orange}{\textit{Community Unit}}, the \textcolor{red}{\textit{Medical Unit}}, and the \textcolor{blue}{\textit{Technological Unit}}.

\subsubsection[]{\includegraphics[width=1em]{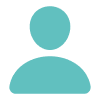} \textcolor{green}{The Role of the Familial Unit}}
In the health information-seeking network, two members of familial unit are central actors: \textit{Patients} themselves and \textit{Patient's Family Members}. We first introduce the roles of each actor and then describe their characteristics and connections between them by cases. 

\leavevmode
\textcolor{green}{\textbf{\textit{The Familial Unit--Patient.}}}
The onset age of HFS patients are around 48.5 ± 14.1 years old \cite{wang1998hemifacial}, ranging from the middle-aged to older adults. Consistent with other studies \cite{mitzner2010older}, we found that some older adult patients are reluctant to use technology, often perceiving it as untrustworthy or unreliable. Consequently, HFS patients in our study have various technology acceptance. We have identified three distinct roles of patients within the network during health information-seeking practices. The first is the \textcolor{green}{\textit{Proactive Information-Seeking Patient}}, who independently gather information from both online and offline sources, often accessing necessary information on their own. Proactive patients may still receive assistance from family members, but primarily in an accompanying and supportive role, as they are generally independent. The second role is the \textcolor{green}{\textit{Low Technological Literacy Patient}}, typically older adults who distrust online information or lack opportunities to use technology due to limited external contact. They usually rely on their children for health information and prefer direct medical consultations. The third role is the \textcolor{green}{\textit{Experienced Patient}}, who has other chronic conditions and a history of hospitalization, making them more attentive to their health status and aware of HFS symptoms earlier, often consulting healthcare professionals directly.

\leavevmode
\textcolor{green}{\textbf{\textit{The Familial Unit--Patient's Family Members.}}}
Family members of HFS patients, either partners or children, assume two roles in the health information-seeking network: \textcolor{green}{\textit{Direct Agent}} and \textcolor{green}{\textit{Companion}}. Direct Agents typically support technology-skeptic patients, acting as mediators who "transform, translate, distort, and modify the meaning of elements they are supposed to carry" \cite{latour2007reassembling}. Since these patients do not seek health information online, their family members assist in gathering both online and offline information. Moreover, in China, accessing healthcare providers and making appointments often involves using mobile phones, a task usually undertaken by family members of technology-skeptic patients. Conversely, Companions align with proactive information-seeking patients, aiding them in information search but not dominating the process. They provide encouragement and support throughout the patients’ journey against the disorder.


P1, a 47-year-old urban patient with 8 years of HFS experience, enacted the \textit{proactive information-seeking patient}. She initially learned that she had HFS while cannot find the cause of it through internet search engines and has tried various treatments, including folk remedy: \textit{"I did as they [colleagues] told me to stick a white paper on my eyelid, some kind of folk remedy, but it didn't work."} Her husband, acting as her companion, assists in inquiring health information within his own social network, as well as providing continuous support from the early stage of illness to post-operative care: \textit{"My husband and my family have been helping me in various ways. I started taking traditional Chinese medicine because my husband's classmate is a doctor at a clinic who specializes in prescribing Chinese herbs."}

On the other hand, patients' children often exhibit greater proficiency with technology and may actively seek health information online on behalf of their parents. For instance, P8, a 70-year-old with 3 years of HFS experience, lives in a rural area and fits the role of a \textit{low technological literacy patient}. She speaks primarily a local dialect, barely speaks Mandarin, and usually interacts only with her husband. P8 uses a basic mobile phone designed for older adults, featuring only a keyboard and screen, to regularly contact her daughter.

Her daughter plays a crucial role as her \textit{direct agent}. P8 does not know which hospital she visits; she trusts and follows wherever her daughter takes her. Consequently, all the health information P8 receives is either directly from her daughter or from medical professionals: \textit{"I know [that I was ill], but we [my husband and I] hadn't gone out [of the rural area before], so we didn't know where to go or what to do. Then, my daughter took me and tried this and that, but nothing seemed to work. My daughter said she heard that [there were good treatments in Beijing], so that's why we went to Beijing."} P8's reliance on her daughter for navigating the healthcare system highlights the significant role family members play in the health information-seeking process, especially for those less familiar with modern technology.

P3, a 70-year-old urban resident with 8 years of HFS experience, exemplifies a patient with \textit{low technological literacy}. She rarely uses a mobile phone and avoids online health information, preferring direct medical consultations. P3 believes that consulting doctors is the most straightforward and effective way to acquire professional medical advice. She views online health information as insufficient and superfluous, considering it limited to symptomatic guidance and often lacking completeness and accuracy. P3 articulated her stance clearly: \begin{quote}
\textit{"I go directly to the hospital when I'm ill and don't search online. I believe consulting a doctor is more direct and helpful. When you're ill, you need to see a doctor. The doctor will tell you how to treat it. Online information might give you some idea about symptoms, but it's often incomplete and unreliable."}
\end{quote}
Her daughter, a physician at a reputable hospital, enacted the role of a \textit{direct agent}, taking responsibility for managing her mother's health conditions and seeking medical advice on her behalf. All the health information P3 received was either directly from her daughter or other medical professionals. Eventually, it was her daughter who introduced her to a leading resource for HFS surgery within the top-tier medical service and persuaded her to undergo the surgery: \textit{"My daughter came across some information suggesting that neurosurgery could be a treatment option. She got this information from the hospital [where my surgery was performed], as she is also a doctor there, so she is quite familiar with hospitals."}

\textit{Experienced patients} are individuals who have already been diagnosed with other health conditions, leading them to be more attentive to any changes in their health status. P11, a 63-year-old patient living in an urban area, has been living with HFS for 15 years. Prior to her HFS diagnosis, she was diagnosed with several other disorders. As a result, she meticulously records her health conditions, taking daily notes to better understand her body: \begin{quote}
    \textit{"After being diagnosed with immune rheumatism, my hospital discharge report revealed that I have rheumatoid arthritis, lupus erythematosus, and Sjögren's syndrome. Consequently, I started keeping detailed records. I jot down notes about my daily medication, any bodily changes, and symptoms I experience. When the eye twitching, characteristic of HFS, began, I included that in my documentation as well."}
\end{quote}
Owing to her diligent record-keeping, she has been able to track her HFS symptoms accurately. This practice aids her in seeking consistent treatment for HFS at the hospital she regularly visits.

\subsubsection[]{\includegraphics[width=1em]{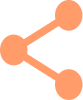} \textcolor{orange}{The Role of the Community Unit}}
Online health information is directly accessible to patients, while offline health information usually requires social networking, either accidental or intentional. We categorize these as the community unit in real life. Patients accessing offline health information outside the medical unit typically engage with two types of actors: those from their \textit{Core Social Circle} and those from the \textit{Peripheral Social Circle} \cite{luo2012neither}.

\leavevmode
\textcolor{orange}{\textbf{\textit{The Community Unit--Core Social Circle.}}}
The core social circle consists of individuals with whom patients have direct contact, including relatives, friends, colleagues, and neighbors. The core social circle plays a crucial role in assisting patients in obtaining health information. Members of this circle are often aware of the patients' situations and may contribute more actively to the patients' health conditions. Within the core social circle, we identified two key roles: the \textcolor{orange}{\textit{First-hand Informant}}, who provides patients with relevant and critical information about HFS, aiding their access to top-tier medical resources, and the \textcolor{orange}{\textit{Second-hand informant}}, transferring essential messages and information from others to HFS patients.

\leavevmode
\textcolor{orange}{\textbf{\textit{The Community Unit--Peripheral Social Circle.}}}
Beyond the core social circle, the peripheral social circle is also a significant source of health information. While this circle may not be the primary source of information for patients, and the connections within it are typically weaker than those in the core social circle, there are instances where patients have encountered or indirectly obtained valuable health information from individuals in this peripheral network. This includes friends of friends, family members of colleagues, and fellow patients met by chance in real life. The roles within this circle are the \textcolor{orange}{\textit{Intermediary}}, who provides significant information indirectly, and the \textcolor{orange}{\textit{Passerby}}, who may not offer particularly useful information, such as fellow patients met in random places.


P4 emphasized the trustworthiness of health information obtained from friends, underscoring the importance of personal connections in the health information-seeking process. P7 shared her experience, highlighting her confidence in health information received from the first-hand informant. She recounted a story about her relatives, which led to her mistrust of local hospitals: \begin{quote}
    \textit{"I don't trust local hospitals because of what happened to my relative. He had a minor surgery that failed, despite the hospital claiming it was simple and he would be discharged in a few days. After that incident, I chose a big hospital in Beijing. Everyone advised against small hospitals for such procedures."}
\end{quote}

On the other hand, P5 acknowledged the influence of colleagues on her treatment choices, despite having reservations: \textit{"I can't even swallow a drop of traditional Chinese medicine, but some of my colleagues are strong proponents of it, so I've tried it before, influenced by their belief"}. These colleagues, acting as the first-hand informants, provided health information that influenced her decisions. P9 discussed how his relative's experience with HFS, who had taken medication without effective results, affected his own approach to treatment, ultimately leading him to forgo drug therapy.

In a different context, P1 detailed how well-informed her friends were about her illness: \textit{"Exactly, they notice when something is wrong because we often meet, and they're all aware of my condition. They know how long I've been dealing with this illness."} In one instance, a friend discovered a similar case in the husband of another friend, providing P1 with invaluable information. Here, P1's friend acted as a second-hand informant, helping P1 acquire critical health information, while the husband of her friend's acquaintance served as an intermediary, offering crucial insights indirectly.

This circumstance is more common among rural patients, who often lack direct access to quality medical resources and rely heavily on word-of-mouth channels. They frequently hear about similar experiences from neighbors or through social connections. D2 noted that information tends to spread quickly in rural communities: \textit{"When rural patients from smaller communities return with information, it tends to spread within about ten days. Once they return to their local area, word spreads, and those with similar conditions come forward"}.

P1 shared an encounter with a passerby while waiting at a bus stop. They had a brief conversation during which the lady revealed that she was convinced that all treatments, including surgery, were ineffective for HFS: \begin{quote}
    \textit{"Once, I met a nearly 70-year-old lady at the bus stop and immediately recognized she had the same condition as me. I asked her, ‘Auntie, do you have Hemifacial Spasm?’ She was surprised and asked how I knew. I explained that I have the same condition and wouldn't have asked otherwise, as it might have seemed impolite. She confirmed she did and mentioned she believed it was incurable. I told her that surgery could help, but she was skeptical. At that time, I didn't try to persuade her further, especially since I hadn't undergone surgery myself. If I met her now, I might tell her that surgery could treat it, as I was also confused about it back then."}
\end{quote}

\subsubsection[]{\includegraphics[width=1em]{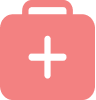} \textcolor{red}{The Role of the Medical Unit}}
Key actors in the medical unit, who significantly influence patients' pathways in seeking medical information and making treatment decisions, include \textcolor{red}{\textit{Specialists}} (e.g., neurosurgeons, neurologists, and ophthalmologists), \textcolor{red}{\textit{Non-Specialized Doctors}} (e.g., general practitioners and traditional Chinese medicine doctors), and \textcolor{red}{\textit{Unlicensed Medical Practitioners}}. We identified three roles they enacted in providing health information based on patients' conditions: \textit{Accurate Information Provider}, \textit{Information-Scant Provider}, and \textit{Misleading Information Provider}.

\leavevmode
\textcolor{red}{\textbf{\textit{The Medical Unit--Accurate Information Provider.}}}
Neurosurgeons enacted the role of Accurate Information Providers, equipped with proper and formal medical training and extensive knowledge of HFS. They specialize in MVD surgery, aimed at providing a complete cure for HFS. When patients consult them, they can offer precise information about HFS and inform the fact that MVD surgery is the only radical cure available. Patients under their care receive an adequate amount of information about HFS and can access treatment or recommended resources directly from them. These neurosurgeons are generally responsible for diagnosing HFS, conducting ward inspections during hospitalization, communicating pre-operative risks, performing the surgery, and conducting follow-up visits as necessary.

\leavevmode
\textcolor{red}{\textbf{\textit{The Medical Unit--Information-Scant Provider.}}}
Ophthalmologists, neurologists, and most general practitioners and traditional Chinese medicine doctors are often the role of information-scant provider in our network. This is because they typically cannot provide patients with comprehensive information about HFS. Some of them may only offer limited messages or treatment options for HFS patients following checkup and inspection, leading us to categorize them under the role of information-scant provider.

Many HFS patients initially struggle to identify the correct clinical department for their specific health issues without proper guidance. This confusion often leads them to inappropriate departments, such as Ophthalmology, under the mistaken belief that eyelid spasms are primarily an eye-related problem. Ophthalmologists, not being specialists in HFS, can only provide limited and often unhelpful information post-examination, as HFS is not an eye disorder. For example, P6 initially believed his eyelid spasms were eye-related and repeatedly consulted ophthalmologists. It was only after learning about HFS from a friend that he realized the issue was not directly related to his eyes. Similarly, P5 and P11, who experienced impaired vision, sought help from ophthalmologists but were left without a clear understanding of their condition despite being told their eyes were in good condition: \textit{"It [HFS] doesn't affect your vision. It twitches, but you can still see things around you. However, over time, I felt that my field of vision became narrower. I even went to see an ophthalmologist and was told there was nothing wrong." (P5)}

D3 emphasized that unless a specialist is involved, other doctors might not recognize the disorder. Although neurologists are educated about HFS, they generally recommend medication and Botox treatments, which are not fully curative and lack long-term efficacy. They often advise patients to opt for conservative treatments, leaving patients without the comprehensive information they seek. D2 explained that non-specialized doctors, such as general practitioners and traditional Chinese medicine practitioners, are usually the first medical providers HFS patients consult when unaware of their condition. However, they often suggest limited treatment options like medication, Chinese herbal remedies, acupuncture, and moxibustion. These treatments are not considered effective cures for HFS, and the lack of information about the more definitive MVD surgery can lead patients to have a constrained mindset when making treatment decisions: 
\begin{quote}
    \textit{"For example, when a patient visits a township clinic, they might not know the appropriate treatment for HFS. Local primary care doctors might either not know about the illness at all or be quite unfamiliar with it. Some might believe Botox is effective, and traditional Chinese medicine practitioners might recommend acupuncture, but they may not realize that surgery is the fundamental cure. Without proper information dissemination, doctors may remain unaware of the illness and its effective treatment." (D2)}
\end{quote}


\leavevmode
\textcolor{red}{\textbf{\textit{The Medical Unit--Misleading Information Provider.}}}
Although some non-specialized doctors can identify and diagnose HFS, it is not a commonly encountered chronic condition for them. As a result, they may lack the necessary knowledge to accurately diagnose HFS, leading to a risk of misdiagnosis and the provision of ineffective treatment options. This can misguide patients with incorrect directions and misinformation, categorizing these doctors as misleading information providers.

While consulting doctors is the most direct method for seeking medical advice, some patients encounter obstacles in accessing appropriate medical services. For HFS patients in rural areas, unlicensed medical practitioners often become an alternative option. These practitioners, without proper medical licensing, typically rely on empirical or practical experiences, claiming their knowledge is inherited from ancestral masters. Their treatments usually include methods like massage and acupuncture. However, rural doctors may lack the necessary expertise in conditions like HFS, often misdiagnosing it as facial paralysis due to its similar facial manifestations. P4, residing in a rural area, explained: \textit{"Rural doctors often treat it [HFS] as facial paralysis because it's more common in our area. They are not familiar with Hemifacial Spasm"}. D3 elaborated that the most significant challenge for rural patients is their limited access to health information about their condition: 
\begin{quote}
    \textit{"The biggest challenge they [rural patients] face is that they may not know what the illness is when they first get it, and some people don't even realize that this [surgery] is the way to cure it until more than ten years after getting the disorder. I think the inability to access information and get timely information is one of the biggest challenges."}
\end{quote}

There are cases of patients turning to barefoot doctors\footnote{Barefoot doctors were unlicensed healthcare practitioners who received basic medical training and worked in rural villages in China.} for treatment. These practitioners commonly misdiagnose HFS as facial paralysis, a more frequent condition in rural areas. For example, P1 consulted several unlicensed healthcare practitioners, including barefoot doctors, who all incorrectly diagnosed her condition as facial paralysis. Despite undergoing their treatments for some time, her condition showed no improvement. She pursued numerous such treatments, driven by a sense of optimism and willingness to spend money on the chance of finding a cure. She described these practitioners as follows: \textit{"He [the unlicensed healthcare practitioner] is Taiwanese, possibly a descendant of some master. In his establishment, there are group photos and other items displayed prominently at the front, showcasing his authority. There are many pictures of him supposedly curing hunchbacks, and he has received numerous banners for these alleged successes."}

\subsubsection[]{\includegraphics[width=1em]{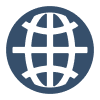} \textcolor{blue}{The Role of the Technology}}
Technology forms our fourth category of actor, serving as an approach for patients to access effective health information online and to locate their ultimate medical resources. This category includes \textcolor{blue}{\textit{Digital Devices}} that facilitate information distribution and transfer (e.g., mobile phones and televisions), as well as the \textcolor{blue}{\textit{Software and Online Services}} these devices encompass (e.g., social media, search engines, digital content featuring videos, pictures, or text). While these technologies are commonplace for many people, it is important to note that the demographic of HFS patients spans from  middle-aged to older adults, and not all of them are adept at using smartphones for information-seeking \cite{yan2020digital}. Consequently, we characterize these technologies as playing two distinct roles based on how patients acquire information: either as an \textit{Active Information Channel} or a \textit{Passive Information Channel}.

\leavevmode
\textcolor{blue}{\textbf{\textit{The Technological Unit--Active Information Channel.}}}
The Internet has become an essential tool for patients seeking health information, offering both convenience and efficiency. Prior research indicates that patients prefer to seek and share health information online using search engines and social media \cite{rice2006influences, fox2011social, fox2013health}. These platforms are considered active channels where patients can proactively find health information. Search engines and social media are the primary channels highlighted in our study. Patients can input keywords into a search engine, yielding a large amount of available information. In China, Baidu is the predominant search engine, while P9 is the exception one claimed that he had only searched on WeChat as it was his most frequently used apps. These platforms serve various purposes, particularly during the awareness of illness stage for self-diagnosis and understanding the causes of their conditions, and later during the treatment decision-making stage to explore potential treatments and collect information about surgeries, healthcare professionals, and hospitals.

Social media platforms, like Tieba and Xiaohongshu, offer forums for individuals with similar health conditions to connect and share experiences. P10, for instance, sought detailed information about surgical procedures on Xiaohongshu. Videos, especially on platforms like Douyin, provide animated demonstrations that help patients understand surgical procedures. However, this approach typically requires some prior knowledge before actively engaging in online searches. Moreover, online information often lacks uniqueness and personalization, leading most patients to combine offline learning with online information to make informed decisions. 

\leavevmode
\textcolor{blue}{\textbf{\textit{The Technological Unit--Passive Information Channel.}}}
We categorize digital content that users encountered by chance or system recommendation as a passive information channel. This often includes content that is not actively sought out but may contain misleading information. For example, P4 came across a video on Douyin recommending a hospital in Shandong for HFS surgery. However, he later consulted with his relatives in Shandong, then he discovered which might be unreliable and even a fraud, as his relatives had never heard about the hospital before: \textit{"I saw on Douyin that there is a hospital in Shandong specializing in HFS surgeries. I look into it and contacted a relative there, but they had never heard of such a hospital and had no information about it."}

Additionally, some patients have gained awareness of their condition through health programs on television. For instance, D2 mentioned that patients have discovered nationally renowned experts through TV program: \textit{"Our director Dr. Liu is a top expert in this surgery nationwide. Back in the early 2000s, he appeared on the CCTV\footnote{China Central Television} health programs 'Health Road' [to promote HFS diagnosis and treatment]."} However, this method of information acquisition has become less common in recent years.

\subsection{Top-Tier Medical Resource Network Strategies}
From the analysis of the connections and interactions among actors in our health information-seeking network, we observed five distinct health information-seeking strategies that ultimately facilitate access to top-tier medical resources. These strategies are named (see Table \ref{tab:strategies}) based on the concepts illustrated in ANT. They include the \textit{Collaborative Inquiry Strategy}, the \textit{Intermediary Strategy}, the \textit{Self-Directed Inquiry Strategy}, the \textit{Observational Strategy}, and the \textit{Utilization of Non-Human Strategy}. Each of these involves a unique combination of units and roles, thereby forming distinct strategic subnetworks within the broader health information-seeking network. 

\subsubsection[]{Network: Collaborative Inquiry Strategy}
The collaborative inquiry strategy involves the \textcolor{green}{\textit{familial unit}} directly networking with top-tier medical resources, allowing for effective and direct access. This strategy typically involves either \textcolor{orange}{\textit{first-hand informants from the community unit}} or the \textcolor{red}{\textit{medical unit}}. Participants can obtain necessary information and establish direct connections with doctors through online channels (e.g., video calls, audio calls, text messages) and offline meetings. This approach facilitates fast and accurate information exchange, enabling immediate responses, making it a rapid and reliable method.

In the familial unit, there may be direct connections with individuals who have recovered from HFS. These individuals, often in good post-operative recovery, provide trustworthy first-hand information. For instance, P12, a 50-year-old proactive information-seeking patient with four years of HFS experience and currently considering MVD surgery, gained insights from a friend who underwent successful surgery. This friend also introduced medical resources to P6, P12's colleague, who similarly had a successful surgery. P12 believes that resources recommended by familiar, recovered individuals are reliable and has been closely monitoring P6's post-operative recovery, which he finds highly informative.

Additionally, the familial unit may have direct connections with medical professionals. Patients themselves or their family members might be doctors, or they may have doctors within their core social circle. For example, P3 is a 70-year-old patient living in Beijing (i.e. first-tier city) with her daughter, benefits from this network. Her daughter, a doctor in a top-tier hospital and acting as a direct agent, leverages her professional network to access medical resources for P3.

\begin{table*}[t]
\centering
\caption{Summary of adopted strategies to acquire the top-tier medical resources.}
\label{tab:strategies}
\Description{Table 2 outlines the five strategies adopted by HFS patients, detailing the typical units involved, the advantages and challenges of each strategy, and their characteristics.}
\setlength\tabcolsep{5pt}
\renewcommand{\arraystretch}{1.2}
\resizebox{\linewidth}{!}{
\begin{tabular}{p{3cm}p{3.5cm}p{3.5cm}p{3.5cm}p{5cm}}
\hline
\textbf{Strategy} &
  \textbf{Typical units involved} &
  \textbf{Advantages} &
  \textbf{Challenges} &
  \textbf{Characteristics} \\ \hline
Collaborative inquiry strategy &
  1. Familial unit \newline 2. Community unit: first-hand informant \newline 3. Medical unit &
  1. Direct networking \newline 
  2. Fast and reliable &
  1. Effective networking required &
  Utilization of online and offline meetings for information exchange, enabling immediate responses. \\ 
\addlinespace
Intermediary strategy &
  1. Familial unit \newline 2. Community unit: second-hand informant and intermediary &
  1. Experienced networking (e.g. recovered patients) \newline 2. No prerequisite &
  1. Indirect sources \newline 2. Time-consuming &
  Reliance on social networks, where information and resources arrive through intermediaries, typically spreads via word-of-mouth and daily conversations. \\ 
\addlinespace
Self-directed inquiry strategy &
  1. Experienced patient \newline 2. Medical unit &
  1. Professional networking \newline 2. Tailored medical advice &
  1. Medical history with doctor consultations &
  Patients can self-refer based on previous medical experiences and directly contact with doctors. \\
 \addlinespace
Utilization of non-human strategy& 1. Familial unit \newline 2. Technology: active information channel & 1. Online networking \newline 2. Convenient access to information & 1. Misinformation \newline 2. Overwhelming similar and limited information \newline 3. Prior knowledge required &Using search engines, browsing social media, and engaging in online communities, often used to enhance and confirm understanding of health information. \\ 
\addlinespace
Observational strategy&
  1. Familial unit \newline 2. Technology: passive information channel &
  1. Unrelated networking \newline 2. Effortless &
  1. Misinformation \newline 2. Serendipitous &
  Involves receiving information passively or accidentally, often through visual media. \\ \hline
\end{tabular}
}
\end{table*}

\begin{figure*}
  \includegraphics[width=\textwidth]{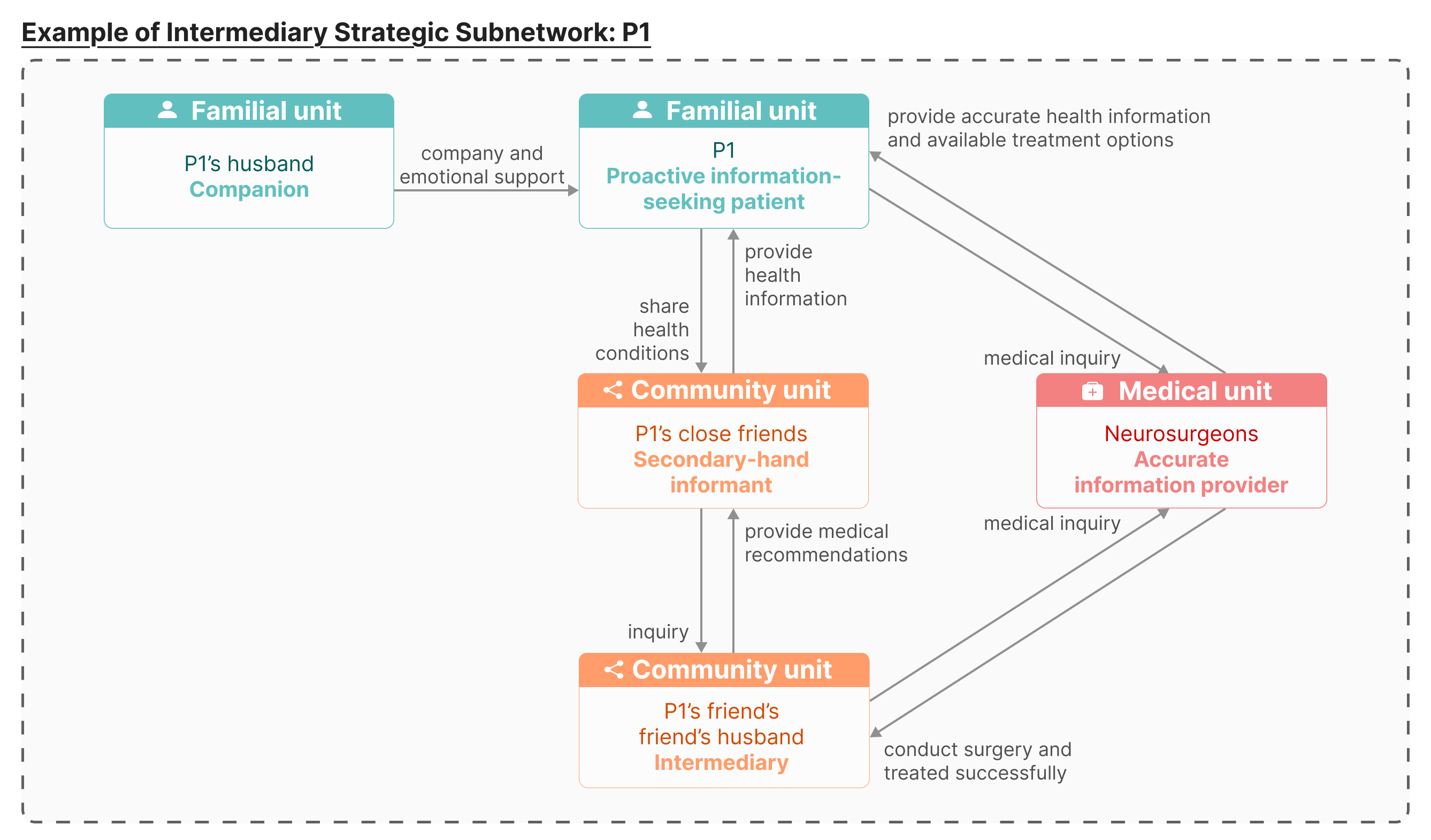}
  \caption{\textbf{Example of Strategic Subnetwork.} P1 was a HFS patient utilizing intermediary strategy to successfully access top-tier medical resources. In the figure, we depict her subnetwork while employing this strategy to illustrate how various actors in different units play their roles to support her journey.}
  \label{fig:ANT-2}
  \Description{Figure 4 illustrates the intermediary strategic subnetwork for P1, involving three main units: the familial unit, community unit, and medical unit. 1. Familial Unit: P1 is a proactive information-seeking patient who actively shares her health conditions and seeks relevant health information. Her husband is identified as a companion, providing emotional support and companionship throughout her journey. 2. Community Unit: P1's close friends serve as secondary-hand informants. P1 shares her health condition with them, and they may provide relevant health information. Additionally, P1's friend's friend's husband acts as an intermediary; having recovered from health issues similar to P1's, he offers medical recommendations to P1's friend. 3. Medical Unit: Neurosurgeons are recognized as accurate information providers. After receiving recommendations from P1's community, she conducts medical inquiries, during which the neurosurgeons offer precise health information and available treatment options tailored to her needs.}
\end{figure*}

\subsubsection[]{Network: Intermediary Strategy}
The intermediary strategy comes into play when direct, effective networking isn't available. In such cases, the \textcolor{green}{\textit{familial unit}} relies on inquiring within their social network, where information and resources often arrive indirectly through intermediaries. This information typically spreads via word-of-mouth and daily conversations, involving the \textcolor{orange}{\textit{second-hand informant and intermediary within the community unit}}. Patients mostly gather information from experienced and recovered patients, introduced by their core social circle.

This indirect approach can be time-consuming, especially since HFS is relatively uncommon in the general population. It is often a lengthy process before patients encounter accurate resources about their condition. This delay is more pronounced for patients in rural areas. D2 highlights the struggles stemming from this delay: 
\textit{"Without proper dissemination of health information, patients might remain oblivious to their condition. Some have suffered for over ten, even twenty years, before seeking appropriate help. This prolonged ignorance can be quite distressing. Many resort to temporary solutions like Botox injections to alleviate symptoms, often repeatedly"}.

Both P1 and P7 discovered top-tier medical resources through conversations within their social network. P1 recounts a friend's awareness of her HFS and their subsequent inquiry about medical resources: \begin{quote}
    \textit{"It was quite a coincidence that a friend brought it up casually. Many of my friends know about my condition and often ask how I'm doing. She had just met a friend whose husband had undergone the surgery. So, she called me, asking about my condition, and I said it was still the same. Then she mentioned that her friend's husband had the surgery about eight years ago and has recovered very well, with no recurrence of the disease at all."}
\end{quote}Similarly, P7 learned about medical resources at a class reunion. There, she met a classmate whose brother is a doctor. She shared a video of her facial symptoms with him for a diagnosis and received contact information for the needed medical resources afterwards.

\subsubsection[]{Network: Self-Directed Inquiry Strategy}
The self-directed inquiry strategy predominantly involves \textcolor{green}{\textit{experienced patients}} and the \textcolor{red}{\textit{medical unit}}. Patients with a history of illnesses tend to be more attentive to their health conditions. Due to their proximity to healthcare providers, they often have regular follow-up visits or maintain contact with doctors who have previously treated them successfully. This access allows them to directly reach out to authoritative and trustworthy doctors for personalized and professional medical advice.

P11, diagnosed with several chronic immune system-related conditions, has extensive experience consulting various doctors and is familiar with the authority of different hospitals. She regularly has follow-up visits with specialists. When she noticed her HFS symptoms, which were different from her usual conditions, she knew exactly which trusted doctors to approach for medical advice. One of these doctors directly connected her with the appropriate medical resources. She explains her choice: \begin{quote}
\textit{"I chose this hospital for my immune disease because I've always trusted it, and the doctor there understands my condition. This doctor referred me to Dr. Liu. I had my child look him up on Baidu, and it turns out he is quite renowned nationwide for performing this type of surgery, with a high success rate. That's why I trust him and have chosen to undergo the surgery at this hospital."}
\end{quote}

\subsubsection[]{Network: Utilization of Non-Human Strategy}
Using search engines, browsing social media, and engaging in online communities are common information-seeking methods today. Most participants generally searched for key symptoms of HFS they experienced to identify potential issues. After learning the name of HFS, they may attempt to search further for potential treatments.
These approaches are convenient but often present challenges in discerning the accuracy of the information due to the vast amount available. As a result, the problem of misinformation persists, and our participants rely on social connections and Guanxi to further validate online information.

Moreover, searching for conditions like HFS, which are not common chronic conditions, the effective information online is limited, and requires a certain degree of digital literacy and prerequisite knowledge. Even with access to professional knowledge, it can be difficult for patients to comprehend such material. Therefore, most patients use this strategy to enhance and confirm their understanding of health information as an assistive strategy. However, it is hard to fully rely on online information for seeking top-tier medical resources for HFS. This strategy usually encompasses the \textcolor{blue}{\textit{active information channels within the technology}} and \textcolor{green}{\textit{familial unit}}.

P6 discovered his disorder through a colleague who had experienced the same issue but had recovered following MVD surgery. He acquired medical information from this colleague and then verified it online to understand the disorder and treatment options in more detail. This additional research helped him decide to undergo the surgery. Similarly, P2 claimed, \textit{"After searching [online], I just thought that I should quickly get the surgery done and not delay it any longer."}. On the other hand, P5 found the top-tier hospital through online information provided by her sister's child: \textit{"My sister's kid did all the research. They looked up which hospital in Beijing is the most authoritative for this condition, and then booked an appointment with that professor for me."}.

\subsubsection[]{Network: Observational Strategy}
The Observational Strategy involves the \textcolor{green}{\textit{familial unit}} receiving information passively or accidentally, often through visual media such as television and social media platforms. This strategy includes the \textcolor{blue}{\textit{passive information channel within the technology unit}}. While encountering this type of information requires minimal effort, it also depends on chance, and there is a risk of encountering misinformation and disinformation. For instance, P4 stumbled upon a video recommending a hospital for HFS surgery, only to discover later that the hospital was not formal and possibly one of the Putian Hospitals\footnote{Putian Hospitals are a group of private hospitals in China, notorious for their aggressive marketing, over-reliance on high-profit medical services, and unethical medical practices.}.

P5, a 54-year-old patient who has been living with HFS for 11 years, has a very close relationship and strong connection with her elder cousin sister. Whenever P5 feels unwell, she turns to her sister, who provides not only emotional support but also practical help, such as assisting in obtaining medications. Her sister, a keen follower of health-related TV programs, diligently takes notes while watching: \textit{"My parents passed away quite early, and it was my sister who took care of me. She loves watching programs on CCTV, and she's very attentive. After watching, she remembers cases, especially common ones like gynecological issues. She takes notes on what medicines to take and what remedies to try. So, whenever I feel unwell, I just give her a call"}. On one occasion, her sister urged P5 to watch a particular program that was discussing symptoms remarkably similar to those P5 was experiencing. It was this program that led P5 to realize that she had been exhibiting symptoms of HFS.


\section{Discussion}
\label{sec:discussion}
In summary, for \textbf{RQ1}, we described HFS patients' health information-seeking network, involving four major units with various combinations of actors and roles as their primary health and healthcare information sources. For \textbf{RQ2}, we identified five strategies HFS patients use to acquire top-tier medical resources, analyzing the pros and cons and the involved actors in each strategy to illustrate the workflows within the network.
In this section, we discussed the similarities and differences in primary health information sources between specialized disorders and other conditions. We highlighted the significance of human infrastructuring in empowering patients access top-tier medical resources and emphasized the potential of the technological unit to help patients overcome challenges and adopt various strategies.

\subsection{Navigating Health Information: Patients' Sources}
Most research on HFS focuses on its clinical, medical, and neurosurgical aspects \cite{aktan2023face, tan2005behind, chaudhry2015hemifacial, rosenstengel2012hemifacial, abbruzzese2011hemifacial}, often neglecting the patients' perspectives. Patients with specialized disorders, part of a marginalized group, have unique challenges and needs that are underrepresented. Our pioneering qualitative study highlights the voices of patients with specialized disorder, revealing their health information-seeking behaviors. We identified four main sources: \textcolor{green}{\textit{familial}}, \textcolor{red}{\textit{medical}}, \textcolor{orange}{\textit{community}}, and \textcolor{blue}{\textit{technological}} units, each offering varying quality of information. This framework aligns with \citet{wilcox2023infrastructuring}'s findings on the infrastructural forms of transgender and non-binary peoples' care, assembled by informal social ecologies, formalized knowledge sources, and self-reflective media, reflecting similarities with our identified units in a similarly marginalized population.

In the familial unit, patients and their family members often function as a cohesive entity in managing health information. Proactive information-seeking patients engage with family members in companion roles, while those with low technological literacy, such as older adults in rural areas, rely heavily on family members as direct agents for finding medical resources. \citet{song2019factors} observed that older adults in rural China depend on proxy health information seeking due to educational and technological limitations. Similarly, our findings reveal that patients struggle to find effective online information due to its repetitive nature, leading to frustration. This mirrors the experiences of individuals with rare diseases, who feel neglected by online and professional medical advice \cite{macleod2017grateful}. \citet{cutilli2010seeking} identified various information sources like TV, radio, newspapers, magazines, the internet, and personal networks that supplement healthcare professional advice. Our study expands on this by showing that patients use online resources and social networks for confirmation and validation. Using ANT, we illustrated the dynamic interplay among these sources, highlighting how patients employ diverse strategies to achieve their health-related goals.

Intermediaries play an important role in reducing information asymmetry of facts and perceptions \cite{yim2017ask}. In HCI, intermediaries are proxy users who bridge the gap between computers and users lacking technological skills or resources, enabling technology usage \cite{sambasivan2010intermediated, kiesler2000troubles}. This phenomenon is particularly prevalent in low-income countries due to economic and cultural factors \cite{parikh2006understanding}. In our findings, direct agents within the familial unit and intermediaries within the community unit are significant sources for improving health information access. They act more as information intermediaries (i.e., infomediaries) rather than intermediaries generally referred to in HCI. 
\citet{parikh2006understanding} proposed the concept of proxy primary and secondary users when designing for intermediated interaction in developing countries, categorizing scenarios into cooperative, dominated, intermediated, and indirect. However, our network is more complex, involving various human and non-human entities. Intermediaries can be constitued by multiple entities, including non-human entities within technological unit \cite{wyatt2008go}. Passive information channels, such as televisions, function as infomediaries by providing random information to patients, while active information channels can mediate health information or connect patients with health providers. For example, patients can access health information through human intermediaries by utilizing technological intermediaries, thereby reducing information asymmetry. For human intermediaries, connections are often facilitated through social networking and guanxi, which we will discuss in Section~\ref{discussion-2}.

\subsection{Accessing Top-Tier Medical Resources: Guanxi and Strong Ties as the Key}
\label{discussion-2}
Specialized disorders require patients to engage in infrastructuring work due to the limited resources and technologies available. Our findings highlight the crucial role of human infrastructures in patient health information-seeking practices. Due to the uncommonness of specialized disorders, technological approaches are less effective, and patients often establish communities through social networks instead. This is particularly evident in the Chinese context, where we observed that most strategies adopted by patients involve social networking and a heavy reliance on guanxi to access top-tier medical resources. In Chinese society, guanxi is a unique cultural element that differs from Western norms. It represents personalized, subjectively close, and resourceful connections between individuals \cite{bian2019guanxi}. The establishment and maintenance of guanxi, typically based on trust and resource exchanges, are essential for acquiring information and reducing uncertainty \cite{fock1998china}. Through guanxi, patients can access scarce information and resources that might otherwise be unavailable \cite{tsui2000guanxi}. This is akin to the increasing significance of guanxi in job acquisition in China, which rose from 46\% in 1978 to 80\% in 2009 \cite{bian2018prevalence}, with strong ties being particularly important \cite{bian1997bringing}.

Social networks play a crucial role in healthcare access for both urban and rural populations \cite{amoah2018social}. In China, strong ties are more significant, and patients trust their familial units for reliable information and resources. This contrasts with Western societies, where \citet{chen2013chinese} found that Chinese individuals rely more on strong ties for job acquisition and promotion, while Western research highlights the advantages of weak ties. A typical example from our study is P3, who leveraged a collaborative inquiry strategy. Her daughter is a doctor in a big city and can quickly connect to medical guanxi and access high-quality medical information and resources. This aligns with \citet{2010Investigation}, who reported that accessing healthcare services through guanxi is a common phenomenon. Moreover, medical guanxi allows patients to obtain high-quality care and services \cite{zhang2021seeking}.
Moreover, we found that rural patients are more likely to use word-of-mouth strategies due to fewer opportunities to connect directly with quality medical resources and lower technological literacy. Thus, they rely more on personal connections, such as close friends, or building new connections through intermediaries.

Although strong ties can significantly help patients connect to quality medical services, patients can still be misdiagnosed and mistreated if they urgently need treatment yet have not found effective connections. Generally, patients regard medical resources in first-tier cities as more reliable and secure, while distrusting rural health systems due to perceived lower quality of medical facilities, services, and doctors' experience. However, in urgent situations, patients may turn to doctors they do not fully trust, such as unlicensed healthcare practitioners, often accessed through recommendations from strong ties, as strong ties provide trust and comfort during uncertainty \cite{krackhardt2003strength}. Therefore, strong ties are crucial in patients' health information-seeking and medical choices, increasing patients' confidence in the rural health system. However, this reliance on strong ties poses challenges to implementing technologies, which we will mention again in Section~\ref{discussion-3}.

\subsection{Empowering Through Technology: Mitigating Challenges in Medical Access}
\label{discussion-3}
Based on our findings, we identified several challenges faced by patients in seeking medical resources, highlighting the potential of \textcolor{blue}{\textit{technology}} to empower them in addressing these issues. 
Firstly, patients who exhibited low health literacy and technological literacy face tough challenges due to disparities in education, geographic isolation, and transportation difficulties. Especially for rural patients who often possess distinct identities and cultural perspectives compared to urban patients \cite{kleinman1978culture}. The lack of healthcare professionals in rural areas can leave these patients feeling powerless in the 'referral games' between rural and urban providers \cite{brundisini2013chronic}. 
For example, both P3 and P8 are low technological patients who rely on their daughters as direct agents. P3, from an urban area, has direct access to top-tier medical services through her daughters. In contrast, P8, from a rural area, and her daughters have struggled to find accurate diagnosis and treatment.
Secondly, patients can be easily misdiagnosed and seek help from unlicensed providers as discussed in Section~\ref{discussion-2}. This situation exemplifies the Chinese adage, "a desperate patient will seek help from any doctor". In such scenarios, the importance of a doctor's credentials lessens, and patients may overlook qualifications due to the urgency for treatment, worsening the asymmetries in doctor-patient relationships. Patients prefer familiar and reliable sources in urgent situations, such as health information from strong ties, believing that technologies may not provide personalized and accurate information. For example, P3 thinks online information only provides symptom introductions and is not accurate enough, leading to reluctance in using online tools.
Thirdly, we observed that patients primarily use technology to enhance and validate their understanding of health information. While non-human strategies can somewhat bridge the gaps and asymmetries in doctor-patient relationships—particularly benefiting rural patients without stringent requirements—these approaches can also lead to misinformation and require prior knowledge. Furthermore, due to the uncommon nature of specialized disorders, the limited and repetitive information available online complicates the access to high-quality information.

In fact, the technological unit holds powerful potential to mitigate these challenges. Information and Communication Technology (ICT) bridges information gaps and addresses power asymmetries by providing previously unavailable information. 
Without ICT, patients often had to rely solely on referrals from local doctors or seek urban doctors without any prior information and assistance. ICT facilitates collaboration between rural and urban doctors, enhances accurate referrals, and fosters digital guanxi between patients and urban doctors, reducing costs and limitations of spreading health information \cite{chib2013enabling}. 
Moreover, ICT allows infomediaries to help patients become more motivated and empowered. Patients can learn more about their disorders and possible future conditions, helping informed decision-making and appropriate actions \cite{khuntia2017patient}. For instance, patients can connect with fellow patients online, leveraging weak ties to share experiences and strategies. \citet{Gage2016Communication} found that patients use online platforms to exchange specialized health information, such as health services, symptom recognition, and treatment protocols. These platforms also facilitate emotional support through empathy, comparing experiences, and offering hope. The internet enables patients to adopt not only self-directed inquiry strategies, with more medical consultation options, especially with specialists, but also collaborative inquiry and intermediary strategies, connecting directly with fellow patients online who can refer them to high-quality medical providers. Therefore, ICT helps infomediaries bring parties together, inform decision-making, and facilitate knowledge exchange, becoming central in the digital ecosystem \cite{yim2017ask}.

To address challenges faced by patients with specific disorders, we propose several future directions. 
First, personalized propagation and education based on disorder characteristics can be beneficial. For instance, HFS patients, typically middle-aged to older adults, can benefit from offline promotion efforts, such as regular free clinic sessions to enhance health and technical literacy. It is essential to boost patients' confidence in and acceptance of technology by promoting the understanding that technologies can serve as valuable sources of health information and provide a means to validate existing information. While concerns about misinformation, fraud, and scams may lead to distrust in technology, these issues can also present in information obtained through social networks.
Second, integrating technology into healthcare policies can provide substantial benefits. The government could provide telemedicine devices to rural patients, enabling remote access to top-tier medical consultations, enhancing health information accuracy, supporting local doctors, and reducing patient costs \cite{stowe2010telecare}. Moreover, incorporating technology into infrastracture can be an effective approach to enhance technology acceptance and literacy. China had nearly 1.1 billion netizens in 2024, with rural netizens accounting for 27.7\%, reaching an Internet penetration rate of 63.8\% in rural areas \cite{CNNIC}. Notably, 36\% of the increased users are people above 50 years old, attracted by the convenience and entertainment of short videos, instant messaging, online payment, and online medical care \cite{CNNIC}.
Third, large language models can serve as virtual doctors, allowing patients to ask medical questions anytime and anywhere. These models offer reliable medical advice and show promising potential for integration into the clinical diagnostic process for rare diseases \cite{Chen2024RareBench}. However, patients may perceive technology as inaccurate and unreliable due to ads and fraud on social platforms, and they are concerned about the impersonal nature of online information. To address these concerns, chatbots can be designed with more personalized, caregiver-like qualities and incorporate voice assistance for more natural interactions, thereby providing a trustworthy image. Moreover, integrating chatbots into online consultations with real doctors may improve patients' confidence in the information provided.
These advancements help mitigate information and power asymmetries in doctor-patient relationships, leading to better health outcomes.


\section{Limitations}
\label{sec:limitations}

This study has several limitations. First, a low response rate led to a small sample size, limiting our ability to comprehensively depict circumstances in both rural and urban China. Second, remote interviews may have hindered rapport-building, potentially restricting insights into personal experiences. Third, the study focused solely on individuals with HFS in China. Future research should expand the participant pool and explore diverse national contexts for more comprehensive and comparative insights.



\section{Conclusion}
\label{sec:conclusion}

In this paper, we explore the health information-seeking practices of HFS patients by interviewing neurosurgeons and patients in China. Using Actor-Network Theory, we analyze the roles, interactions and workflows of both human and non-human entities within the health information-seeking network. We identify five strategies for accessing top-tier medical resources, detailing their pros and cons. We disucuss how patients navigate health information, highlighting the empowerment through guanxi, strong ties, and the role of intermediaries. Despite the significant role of human infrastructure, the journey for patients is often long, difficult, and complex, involving numerous attempts to find adequate medical care. We propose implications for technologies to mitigate these challenges and empower patients.
Our goal is to enhance research on patient engagement and to understand how healthcare practitioners, designers, developers, and government can improve health awareness and information acquisition. Ultimately, we aim to help individuals with specialized disorder overcome information and power asymmetries in doctor-patient relationships, gaining better access to medical services and treatment options.


\begin{acks}
This work is supported by the Natural Science Foundation of China under Grant No. 62472244 and 62132010, the Beijing National Research Center for Information Science and Technology (BNRist), the Tsinghua University Initiative Scientifc Research Program, and the Undergraduate Education Innovation Grants, Tsinghua University.
We thank all participants for taking part in our study, Bo Hei for assisting with participant recruitment. We also extend our gratitude to all reviewers, past and present, for their valuable comments that significantly improved this paper.
\end{acks}

\bibliographystyle{ACM-Reference-Format}
\bibliography{sample-base}

\end{document}